\documentclass[12pt,a4paper]{article}
\usepackage[T1]{fontenc}        
\usepackage[dvips]{epsfig}
\title{
  {\vspace{-2cm} \normalsize
     \epsfig{figure=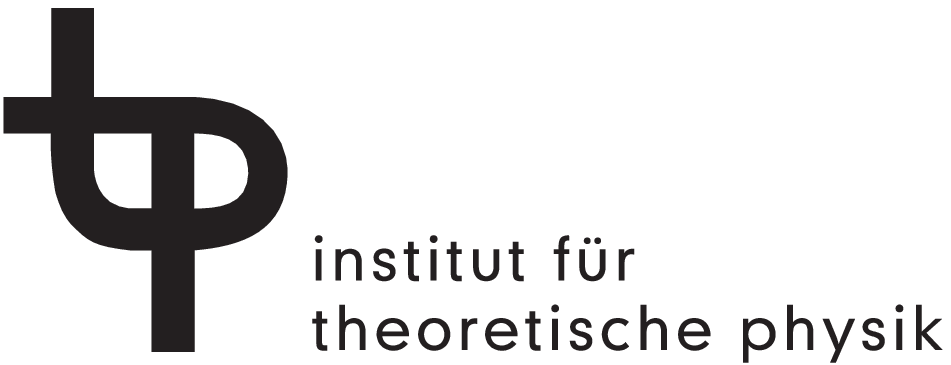,width=80mm}
     \hfill\parbox[b][30mm][t]{35mm}{MS-TP-03-27 \\
                                     hep-lat/0311032}  }\\[25mm]
  Chiral perturbation theory\\
  for lattice QCD with a twisted mass term
  }
\author{Gernot M\"unster and Christian Schmidt\\
        Institut f\"ur Theoretische Physik,
        Universit\"at M\"unster\\
        Wilhelm-Klemm-Str.~9, D-48149 M\"unster, Germany\\
        e-mail: munsteg@uni-muenster.de, sschmidt@uni-muenster.de}
\date{November 21, 2003\\(revised: February 26, 2004)}
\newcommand{\I}{\ensuremath{\mathrm{i}\,}}
\newcommand{\E}{\ensuremath{\mathrm{e}\,}}
\begin{document}
\maketitle

\begin{abstract}
Quantum Chromodynamics on a lattice with Wilson fermions and a chirally
twisted mass term for two degenerate quark flavours is considered in the
framework of chiral perturbation theory. The pion mass and decay constant
are calculated in next-to-leading order including terms linear in the
lattice spacing $a$.
\end{abstract}
%
Numerical simulations of Quantum Chromodynamics with dynamical quarks are
facing a difficulty with light quark masses. Due to a severe slowing down of
the common algorithms for small quark masses it is presently not possible to
implement realistic values for the masses of the u- and d-quarks. In this
context chiral perturbation theory \cite{Weinberg,GL1,GL2} is a very useful
tool. It amounts to an expansion around the chiral limit where the quarks
are massless. Meson masses and other physical quantities are then expanded
in powers of quark masses (modified by logarithms). In its range of
applicability, chiral perturbation theory can be used to extrapolate the
results of Monte Carlo simulations of QCD into the region of small quark
masses. On the other hand, numerical simulations of lattice QCD can provide
information about low energy parameters of chiral perturbation theory, the
\textit{Gasser-Leutwyler coefficients}, see \cite{Wittig} for a review.

When chiral perturbation theory is applied to lattice QCD, the lattice
spacing $a$ enters as a second expansion parameter. Correspondingly, the low
energy effective Lagrangean contains additional terms proportional to powers
of $a$, representing the lattice artifacts \cite{Rupak-Shoresh}. Physical
quantities appear in double expansions in quark masses and in $a$. A
comparison of numerical results with chiral perturbation theory in
next-to-leading order has been made in \cite{qqq1,qqq2}.

An attractive framework for numerical simulations of QCD with Wilson
fermions is the lattice formulation with a chirally twisted quark mass
matrix \cite{TM,Frezzotti}. It allows to avoid problems with quark zero
modes and simplifies the renormalization procedure. In view of Monte Carlo
calculations in twisted mass lattice QCD it is desirable to extend chiral
perturbation theory to this case.

In this letter we consider lattice QCD with $N_f =2$ quark flavours and a
twisted mass term. For simplicity we restrict ourselves to the case of
degenerate quark masses ($m_u=m_d=m$). The pion mass and the pion decay
constant are calculated in chiral perturbation theory in next-to-leading
(one-loop) order, including lattice terms linear in the lattice spacing $a$.
A new feature of the twisted mass case is the fact that the minimum of the
effective Lagrangean is shifted by an amount proportional to $a$ and the
expansion has to be rearranged correspondingly.

In lattice QCD a chirally twisted mass term can be introduced in the form
$\bar{q} \, M(\omega) q$ with
\begin{equation}
M(\omega) = m' \, \E^{\I \omega \gamma_5 \tau_3}
= m' \cos (\omega) \, {\bf 1} + \I m' \sin (\omega) \,
\gamma_5 \tau_3
\equiv m + \I \mu \gamma_5 \tau_3\,,
\end{equation}
where
\begin{equation}
m = m' \cos( \omega)\,, \quad \mu= m' \sin( \omega)\,,
\end{equation}
and $\tau_b$ are Pauli matrices. In the continuum the chiral rotation of the
mass term can be removed by a chiral transformation of the quark fields
according to
\begin{equation}
q = \E^{-\I \omega \gamma_5 \tau_3/2}\, q'\,,
\end{equation}
which leaves the kinetic term invariant. On the lattice, however, it is not
possible to remove the chiral twist due to the presence of lattice
artifacts.

Chiral perturbation theory is based on the low energy effective Lagrangean,
which describes the dynamics of pions in terms of SU(2)-valued matrices
\begin{equation}
U(x)=\exp \left( \frac{\I}{F_0} \, \pi_b(x) \tau_b \right),
\end{equation}
where $\pi_b(x)$ are the pion fields. In the continuum it is given to
leading order by
\begin{equation}
\mathcal{L}_2 = \mathcal{L}_{2,kin} + \mathcal{L}_m
= \frac{F_0^2}{4}\,\mbox{Tr} \left( \partial_\mu U^\dagger \, \partial^\mu U 
\right)
- \frac{F_0^2}{4}\,\mbox{Tr} \left( \chi U^\dagger + U \chi^\dagger
\right). 
\end{equation}
The chiral symmetry breaking mass term contains the matrix
\begin{equation}
\chi= 2 B_0 M = 2 B_0 m \, {\bf 1} \equiv \chi_0 \, {\bf 1},
\end{equation}
with
\begin{equation}
B_0 = \frac{1}{F_0^2} \, \left< 0 | \bar{u} u | 0 \right>
= \frac{1}{F_0^2} \, \left< 0 | \bar{d} d | 0 \right>.
\end{equation}
To lowest order in $m$ the pion mass is given by
\begin{equation}
m_\pi^2= \chi_0 = 2 B_0 m.
\end{equation}

Lattice QCD can be treated in chiral perturbation theory by adding terms
which represent the lattice artifacts. The chiral Lagrangean to lowest order
is extended to
\begin{equation}
\mathcal{L}_2= \mathcal{L}_{2,kin}+\mathcal{L}_m 
- \frac{F_0^2}{4}\,\mbox{Tr} \left( \rho\, U^\dagger + U  \rho^\dagger
\right),
\end{equation}
where
\begin{equation}
\rho= 2 W_0 a \, {\bf 1}
\end{equation}
and the coefficient $W_0$ parametrizes the lattice artifacts.

Now let us consider the effective Lagrangean for QCD with a twisted mass
term. The twisting of the mass term
\begin{equation}
\chi \to \chi( \omega)
= 2 B_0 \, \E^{-\I \omega \tau_3/2} \, M \, \E^{-\I \omega \tau_3/2}
= 2 B_0 m \, \E^{-\I \omega \tau_3}
\end{equation}
can be undone in the kinetic and mass terms by a chiral transformation of
the axial type
\begin{equation}
U = \E^{-\I \omega \tau_3/2} \, U' \, \E^{-\I \omega \tau_3/2}.
\end{equation}
The lattice term, however, is transformed to
\begin{equation}
\mbox{Tr} \left( \rho(\omega) \, U'^\dagger + U' \, \rho(\omega)^\dagger
\right)
\end{equation}
with the matrix
\begin{equation}
\rho(\omega) = \E^{\I \omega \tau_3} \rho\,.
\end{equation}  
In this way the twist can be transferred to the lattice term. Omitting the
primes on the chiral field $U$, the leading order effective Lagrangean reads
\begin{equation}
\mathcal{L}_2 
= \frac{F_0^2}{4}\,\mbox{Tr} \left( \partial_\mu U^\dagger\, \partial^\mu U
\right)
- \frac{F_0^2}{4}\,\mbox{Tr} \left( (\chi+\rho(\omega))\, U^\dagger
+ U (\chi+\rho(\omega))^\dagger \right)
\end{equation}
with
\begin{equation}
\chi= 2 B_0 m' \, {\bf 1}.
\end{equation}
In next-to-leading order the following terms contribute
\cite{Rupak-Shoresh}:
\begin{eqnarray}
\mathcal{L}_4 & = & \frac{F_0^2}{4}\,\mbox{Tr} \left( \partial_\mu U
\partial_\mu U^\dagger \right)
- \frac{F_0^2}{4}\,\mbox{Tr} \left( \chi U^\dagger + U \chi^\dagger \right)
- \frac{F_0^2}{4}\,\mbox{Tr} \left( \rho(\omega) U^\dagger
+ U \rho(\omega)^\dagger \right) \\
& & - L_1 \left[ \mbox{Tr} \left( \partial_\mu U \partial_\mu U^\dagger
\right) \right]^2
- L_2\, \mbox{Tr} \left( \partial_\mu U \partial_\nu U^\dagger \right) 
\mbox{Tr} \left( \partial_\mu U \partial_\nu U^\dagger \right) \nonumber \\
& & - L_3\, \mbox{Tr} \left( \left[ \partial_\mu U \partial_\mu  U^\dagger
\right]^2 \right)
+ L_4\, \mbox{Tr} \left( \partial_\mu U \partial_\mu  U^\dagger \right)
\mbox{Tr} \left( \chi U^\dagger + U \chi^\dagger \right) \nonumber \\
& & +  W_4\, \mbox{Tr} \left( \partial_\mu U \partial_\mu  U^\dagger \right)
\mbox{Tr} \left( \rho(\omega) U^\dagger + U \rho(\omega)^\dagger \right)
+ L_5\, \mbox{Tr} \left( \partial_\mu U \partial_\mu  U^\dagger
\left[ \chi U^\dagger + U \chi^\dagger \right] \right) \nonumber \\
& & + W_5\, \mbox{Tr} \left( \partial_\mu U \partial_\mu  U^\dagger
\left[ \rho(\omega) U^\dagger + U \rho(\omega)^\dagger \right] \right)
- L_6 \left[ \mbox{Tr} \left( \chi U^\dagger + U \chi^\dagger \right)
\right]^2 \nonumber \\
& & - W_6\, \mbox{Tr} \left( \chi U^\dagger + U \chi^\dagger \right)
\mbox{Tr} \left( \rho(\omega) U^\dagger + U \rho(\omega)^\dagger \right)
- L_7 \left[ \mbox{Tr} \left( \chi U^\dagger - U \chi^\dagger \right)
\right]^2 \nonumber \\
& & - W_7\, \mbox{Tr} \left( \chi U^\dagger - U \chi^\dagger \right)
\mbox{Tr} \left( \rho(\omega) U^\dagger - U \rho(\omega)^\dagger \right)
- L_8\, \mbox{Tr} \left( \chi U^\dagger \chi U^\dagger
+ U \chi^\dagger U \chi^\dagger \right) \nonumber \\
& & - W_8\, \mbox{Tr} \left( \chi U^\dagger \rho(\omega) U^\dagger
+ U \rho(\omega)^\dagger U \chi^\dagger \right)
+ \mathcal{O} (a^2). \nonumber 
\end{eqnarray}
Spurion analysis shows that the twisting of the mass term does not produce
further terms in $\mathcal{L}_4$, but amounts to replacing $\rho$ by
$\rho(\omega)$.

In contrast to the untwisted case, the minimum of the effective action is
not located at vanishing fields $\pi_b = 0$, corresponding to
$U=\mathbf{1}$, but at the point
\begin{equation}
\tilde{\pi}_3 = F_0 \frac{W_0 a}{B_0 m'} \, \sin \omega
\left[ 1 - \frac{8 \chi_0}{F_0^2} (4 L_6 + 2 L_8 - 2 W_6 - W_8 ) \right]
+ \mathcal{O} (a^2),
\quad \tilde{\pi}_1 = 0, \quad \tilde{\pi}_2 = 0.
\end{equation}
Therefore a shift
\begin{equation}
\pi_3 = \pi'_3 + \tilde{\pi}_3 
\end{equation}
has to be performed before expanding the effective action.

Using the resulting expression we calculated the pion propagator in
next-to-leading order. From it the pion mass and wave function
renormalization are obtained. For the pion mass we get
\begin{eqnarray}
m_{\pi}^2 & = & \chi_0 + \rho_0 + 8 \frac{\chi_0^2}{F_0^2}
(4 L_6^r + 2 L_8^r - 2 L_4^r - L_5^r) \nonumber \\
& & +8 \frac{\chi_0 \rho_0}{F_0^2}
(4 W_6^r + 2 W_8^r - 2 W_4^r - W_5^r - 2 L_4^r - L_5^r)\nonumber \\
& & + \frac{(\chi_0+\rho_0)^2}{32 \pi^2 F_0^2}\,
\ln \left(\frac{\chi_0 + \rho_0}{\Lambda^2} \right),
\end{eqnarray}
where 
\begin{equation}
\chi_0 = 2 B_0 m'\,, \quad \rho_0 = 2 W_0 a \cos \omega\,,
\end{equation}
$L_k^r$ are the renormalized chiral parameters and $\Lambda$ is the
renormalization scale. The wavefunction renormalization in dimensional
regularization is given by
\begin{eqnarray}
Z &=& 1 - \frac{8}{F_0^2} \chi_0 (2 L_4 + L_5)
- \frac{8}{F_0^2} \rho_0\, (2 W_4 + W_5)\nonumber\\ 
&&- \frac{1}{24 \pi^2 F_0^2} (\chi_0 + \rho_0)
\left[ \frac{2}{\epsilon} + \ln (4\pi) - \gamma +1
- \ln \left( \frac{\chi_0 + \rho_0}{\Lambda^2} \right) \right].
\end{eqnarray}
Setting the twist angle $\omega$ to zero, the expression for $m_{\pi}^2$ is
consistent with the result of \cite{Rupak-Shoresh}. Apart from numerical
factors which depend on $N_f$, to this order of chiral perturbation theory
the chiral twist amounts to adding the factor $\cos \omega$ to $\rho$. It
should be noted, however, that this rule will not apply to higher orders,
where the above mentioned shift in the pion fields introduces new vertices.

In the case of maximal twist, $\omega = \pi/2$, the lattice artifacts vanish
to the order of the expansion considered here, as has been observed for
lattice QCD in general in \cite{FR1,FR2}.

The second physical quantity we calculated is the pion decay constant
$F_{\pi}$, given by
\begin{equation}
\langle 0 | J_A^{\mu,a} | \pi_b(p) \rangle = \I F_{\pi} p^\mu \delta_{ab}\,,
\end{equation}
where $J_A$ is the axial current. The Noether procedure yields an expression
for the axial current, which depends on the coefficients $L_k$ and $W_k$ and
contains products of the pion fields. A one-loop calculation gives
\begin{equation}
F_{\pi} = F_0 \left( 1 + \frac{4}{F_0^2} [  \chi_0 \, (2 L^r_4 + L^r_5)
+ \rho_0 \, (2 W_4^r + W_5^r)] - \frac{1}{16\pi^2 F_0^2}
\left[ (\chi_0 + \rho_0) \ln \frac{\chi_0 + \rho_0}{\Lambda^2} \right]
\right).
\end{equation}
In the limit $a \to 0$ this coincides with the result of \cite{GL2}.

The formulae given here can be used in the analysis of numerical results
from unquenched simulations of twisted mass lattice QCD with two light
flavours.  A useful technique in numerical simulations of QCD is partial
quenching of quarks. In view of applications to partially quenched
simulations it suggests itself to generalize the results above to partially
quenched chiral perturbation theory \cite{MS}.

We thank I.~Montvay, E.~Scholz and F.~Farchioni for discussions.

%
\end{document}